\documentclass{Rinton-P9x6}

\begin{document}
\def\IJMPA #1 #2 #3 {{\sl Int.~J.~Mod.~Phys.}~{\bf A#1}\ (19#2) #3$\,$}
\def\MPLA #1 #2 #3 {{\sl Mod.~Phys.~Lett.}~{\bf A#1}\ (19#2) #3$\,$}
\def\NPB #1 #2 #3 {{\sl Nucl.~Phys.}~{\bf B#1}\ (19#2) #3$\,$}
\def\PLB #1 #2 #3 {{\sl Phys.~Lett.}~{\bf B#1}\ (19#2) #3$\,$}
\def\PR #1 #2 #3 {{\sl Phys.~Rep.}~{\bf#1}\ (19#2) #3$\,$}
\def\JHEP #1 #2 #3 {{\sl JHEP}~{\bf #1}~(19#2)~#3$\,$}
\def\PRD #1 #2 #3 {{\sl Phys.~Rev.}~{\bf D#1}\ (19#2) #3$\,$}
\def\PTP #1 #2 #3 {{\sl Prog.~Theor.~Phys.}~{\bf #1}\ (19#2) #3$\,$}
\def\PRL #1 #2 #3 {{\sl Phys.~Rev.~Lett.}~{\bf#1}\ (19#2) #3$\,$}
\def\RMP #1 #2 #3 {{\sl Rev.~Mod.~Phys.}~{\bf#1}\ (19#2) #3$\,$}
\def\ZPC #1 #2 #3 {{\sl Z.~Phys.}~{\bf C#1}\ (19#2) #3$\,$}
\def\PPNP#1 #2 #3 {{\sl Prog. Part. Nucl. Phys. }{\bf #1} (#2) #3$\,$}
\title{Massive neutrinos and lepton-flavour-violating processes}

\author{J.A. Casas}

\address{Instituto de Estructura de la Materia, CSIC\\
Serrano 123, 28006 Madrid}

\author{A. Ibarra}

\address{Department of Physics, Theoretical
Physics, University of Oxford \\
1 Keble Road, Oxford OX1 3NP,
United Kingdom}  


\maketitle

\abstracts{
If neutrino masses and mixings are suitable to explain the atmospheric
and solar neutrino fluxes, this amounts to contributions to FCNC
processes, in particular  $\mu\rightarrow e, \gamma$. If the theory
is supersymmetric and the origin of the masses is a see-saw mechanism,
we show that the prediction for  BR($\mu\rightarrow e, \gamma$) is in
general larger than the experimental upper bound, especially if the
solar data are explained by a large angle MSW effect, which recent
analyses suggest as the preferred scenario.}

\section{See-saw, RG-induced LFV soft terms and $l_i\rightarrow l_j, \gamma$}

In the pure Standard Model, flavour is exactly conserved in the
leptonic sector since one can always choose a basis in which the
(charged) lepton Yukawa matrix, $\bf{Y_e}$, and gauge interactions are
flavour-diagonal. If neutrinos are massive and mixed, as suggested by
the observation of atmospheric and solar fluxes
\cite{SK}, this is no longer
true and there exists a source of lepton flavour violation (LFV), 
in analogy with the
Kobayashi--Maskawa mechanism in the quark sector.  Unfortunately, due
to the smallness of the neutrinos masses, the predicted branching
ratios for these processes are so tiny that they are completely
unobservable, namely  BR$(\mu \rightarrow e \gamma) < 10^{-50}$
\cite{meg_SM}.

In a supersymmetric (SUSY) framework the situation is completely different.
Besides the previous mechanism, supersymmetry provides new direct
sources of flavour violation in the leptonic sector, namely the
possible presence of off-diagonal soft terms \cite{masiero}.
In a
self-explanatory notation, they have the form
\bea  -{\cal L}_{\rm soft}=\left(m_L^2\right)_{ij} \bar L_iL_j+
\left(m_{e_R}^2\right)_{ij} {\bar e}_{Ri} e_{Rj} + \left({A_e}_{ij}
{e_R^c}_i H_1 L_j +{\rm h.c.}\right) + {\rm etc.}\; ,  \eea
where we have written explicitly just the soft breaking terms in the
leptonic sector, namely scalar masses and trilinear scalar terms. All
the fields in the previous equation denote just the corresponding
scalar components. Concerning flavour
violation the most conservative starting point for ${\cal L}_{\rm
soft}$ is the  assumption of universality, which corresponds to take
\bea
\label{universal} 
\left(m_L^2\right)_{ij} = m_0^2\ {\bf 1},\;\;\;
\left(m_{e_R}^2\right)_{ij} = m_0^2\ {\bf 1},\;\;\; {A_e}_{ij}=A_0\
{\bf Y_e}_{ij}\;,  \eea
so that working in the $L_i$ and ${e_R}_i$ basis where  ${\bf Y_e}$ is
diagonal, the soft terms do not contain off-diagonal (lepton flavour
violating) entries.

%
%
%

It turns out, however, that even under this extremely conservative
assumption, if neutrinos are massive, radiative corrections may
generate off-diagonal soft terms.

The most interesting example of this occurs when neutrino masses are
produced by a (supersymmetric) see-saw mechanism \cite{seesaw}. 
This is based upon a superpotential
\bea
\label{superp}
W=W_{0} - \frac{1}{2}{\nu_R^c}^T{\cal M}\nu_R^c +{\nu_R^c}^T {\bf
Y_\nu} L\cdot H_2, \eea
where $W_{0}$ is the observable superpotential, except for neutrino masses, 
of the preferred version of the supersymmetric SM, e.g. the MSSM. The extra 
terms involve three additional neutrino chiral fields (one per generation;
indices are suppressed) not charged under the SM group: ${\nu_{R}}_i$
($i=e,\mu,\tau$). ${\bf Y_\nu}$ is the matrix of neutrino Yukawa
couplings,  $L_i$ ($i=e,\mu,\tau$) are the  left-handed lepton
doublets and $H_2$ is the hypercharge $+1/2$ Higgs doublet. The Dirac
mass matrix is given by ${\bf m_D}={\bf Y_\nu}\langle H_2^0\rangle$.
Finally, ${\cal M}$ is a $3\times 3$ Majorana mass matrix 
whose natural scale, say $M$, is much
larger than the electroweak scale or any soft mass. 
%
%
Below $M$ the
theory is  governed by an effective superpotential
$W_{eff}=W_{0}+\frac{1}{2}({\bf Y_\nu}L\cdot H_2)^T{\cal
M}^{-1}({\bf Y_\nu}L\cdot H_2)$, 
%
%
%
%
obtained by integrating out the heavy neutrino fields in
(\ref{superp}). Hence, the effective neutrino mass matrix, 
${\cal M}_\nu$, is given by
\bea
\label{kappa}
\kappa \equiv {\cal M}_\nu/ \langle H_2^0\rangle^2= {\bf Y_\nu}^T {\cal
M}^{-1} {\bf Y_\nu},  \eea
where $\langle H_2^0\rangle^2=v_2^2=v^2 \sin^2\beta$ and   $v=174$
GeV. The experimental data about neutrino masses and mixings are
referred to the ${\cal M}_\nu$ matrix, or equivalently $\kappa$,
evaluated at low energy (electroweak scale)\footnote{It should
be noted that eq.(\ref{kappa}) is defined at  the ``Majorana scale'', $M$.
Therefore, in order to
compare to the experiment one has still to run $\kappa$ down to low
energy through the corresponding RGE.}.

Turning back to the structure of the SUSY soft-breaking terms, the
universality condition (\ref{universal}) can only be imposed at a
certain scale, typically at the scale at which the soft breaking terms
are generated, e.g. $M_X$ in GUT models. Below that scale,
the RGEs of the soft terms, which contain non-diagonal contributions
proportional to  ${\bf Y_\nu^+}{\bf Y_\nu}$, induce off-diagonal soft
terms \cite{Borzumati:1986qx,topdown,Hisano:1996cp,Hisano:1999fj}
These contributions are decoupled at the characteristic scale
of the right-handed neutrinos, $M$.  
More precisely,  in
the leading-log approximation
\footnote{We use the leading-log
approximation through the text in order to make the results easily
understandable. Nevertheless, the numerical results, to be exposed below,
have been obtained by integrating the full set of RGEs.}
, the
off-diagonal  soft terms at low-energy are given by
\bea
\label{softafterRG} 
\left(m_L^2\right)_{ij} & \simeq & \frac{-1}{8\pi^2}(3m_0^2 + A_0^2)
({\bf Y_\nu^+}{\bf Y_\nu})_{ij} \log\frac{M_X}{M}\ , \nonumber\\
\left(m_{e_R}^2\right)_{ij} & \simeq & 0\ , \nonumber\\  ({A_e})_{ij}&
\simeq &   \frac{-3}{8\pi^2} A_0 Y_{l_i} ({\bf Y_\nu^+}{\bf
Y_\nu})_{ij} \log\frac{M_X}{M}\;,   \eea
where $i\neq j$ and $Y_{l_i}$ is the Yukawa coupling of the  charged
lepton $l_i$.

The previous off-diagonal soft terms induce LFV processes, like
$l_i\rightarrow l_j, \gamma$.
The precise form of  BR($l_i\rightarrow
l_j, \gamma$) that we have used in our computations is a rather
cumbersome expression ~\cite{Hisano:1996cp}. However, for the sake
of the physical discussion it is interesting to think in the 
mass-insertion approximation to identify the dominant contributions. As
discussed in ref. \cite{Hisano:1999fj}, these correspond to the mass-insertion
diagrams enhanced by $\tan\beta$ factors. All of them are proportional
to ${m_L^2}_{ij}$, and have the generic form shown in Fig.~1.  
Thus the size of the braching ratios is given by

\begin{figure}
\centerline{\hbox{
\psfig{figure=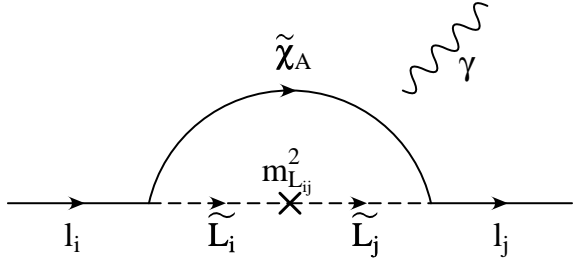,width=7.5cm}}}
\caption
{\footnotesize   Dominant 
Feynman diagrams contributing to 
$l_i\rightarrow l_j, \gamma$ in the mass-insertion approximation.
$\tilde L_i$ are the slepton doublets in the basis where 
the gauge interactions and the charged-lepton Yukawa couplings are 
flavour-diagonal. $\tilde \chi_A$ denote the charginos and neutralinos.
}
\end{figure}

%
\bea
\label{BR2}
\hspace{-0.5cm}
{\mathrm{BR}}(l_i\rightarrow l_j, \gamma) &\sim& \frac{\alpha^3}{G_F^2}
\frac{{|m_L^2}_{ij}|^2}{m_S^8}\tan^2\beta 
\nonumber\\
&\sim&
\frac{\alpha^3}{G_F^2m_S^8}
\left|\frac{-1}{8\pi^2}(3m_0^2+A_0^2)\log\frac{M_X}{M}\right|^2
\left|({\bf Y_\nu^+}{\bf Y_\nu})_{ij}\right|^2\tan^2\beta 
\eea
where we have used  eqs.(\ref{softafterRG}).
The ${\bf Y_\nu^+}{\bf Y_\nu}$ matrix is therefore the crucial
quantity for the computation of
BR($l_i\rightarrow l_j, \gamma$).
Hence, in order to make predictions on BR($l_i\rightarrow l_j, \gamma$) 
we need to determine the most
general form of  ${\bf Y_\nu}$ and ${\bf Y_\nu^+}{\bf Y_\nu}$,
compatible with all  the phenomenological requirements.
Recall that the latter are referred to the ${\cal M}_\nu$ matrix,
evaluated at low energy, rather than to  ${\bf Y_\nu}$ itself.
So, this is a non-trivial task that we discuss in the next section.
Notice also the  strong dependence of BR($l_i\rightarrow l_j, \gamma$)
on $\tan \beta$ and the fact that the larger (smaller) the initial scale 
at which universality is imposed, the larger 
BR($l_i\rightarrow l_j, \gamma$).

\section{General textures reproducing experimental data}

Working in the flavour basis in which the charged-lepton Yukawa
matrix, $\bf{Y_e}$, and gauge interactions are flavour-diagonal,  
the neutrino mass matrix, 
${\cal M}_\nu$, or equivalently the $\kappa$ matrix defined in 
eq.(\ref{kappa}), 
is diagonalized by the MNS \cite{MNS} matrix $U$ according to
\be
\label{Udiag}
U^T{\kappa } U=\mathrm{diag}(\kappa_1,\kappa_2,\kappa_3)\equiv
D_\kappa, \ee
where $U$ is a unitary matrix that relates  flavour to mass eigenstates.
It is possible, and sometimes convenient, to choose  $\kappa_i\geq
0$. Then, $U$ can be written as $U=V\cdot 
\mathrm{diag}(e^{-i\phi/2},e^{-i\phi'/2},1)$, 
where $\phi$ and $\phi'$ are CP violating phases (if different from
$0$ or $\pi$) and $V$ has the ordinary form of a CKM matrix
\be \label{Vdef} V=\pmatrix{c_{13}c_{12} & c_{13}s_{12} & s_{13}e^{-i\delta}\cr
-c_{23}s_{12}-s_{23}s_{13}c_{12}e^{i\delta} & c_{23}c_{12}-s_{23}s_{13}s_{12}e^{i\delta} & s_{23}c_{13}\cr
s_{23}s_{12}-c_{23}s_{13}c_{12}e^{i\delta} & -s_{23}c_{12}-c_{23}s_{13}s_{12}e^{i\delta} &
c_{23}c_{13}\cr}.  \ee

The experimental information about neutrinos 
consists of information about
the low-energy spectrum of neutrinos, contained in
$D_{\kappa}$, and about the neutrino mixing angles (and CP phases),
contained in $U$. Let us discuss them in order.

The experimental (solar and atmospheric) data \cite{SK} strongly suggest a
hierarchy of neutrino mass-splittings, $\Delta \kappa_{sol}^2\ll
\Delta \kappa_{atm}^2$. Numerically, $\Delta \kappa_{atm}^2 \sim
3\times 10^{-3} \mathrm{eV}^2/v_2^4$ \cite{Gonzalez-Garcia:2001sq},
while the value
of $\Delta \kappa_{sol}^2$ depends on the solution considered to
explain the solar neutrino problem \cite{Pontecorvo,MSW}, i.e.
large-angle, small angle or LOW  MSW solutions
(LAMSW, SAMSW and LOW respectively), or vacuum oscillations solution (VO).
They
require, in $\mathrm{eV}^2/v_2^4$ units,  $\Delta \kappa_{sol}^2\sim
3 \times10^{-5}$,
$10^{-7}$, $5\times 10^{-6}$ and  $8\times 10^{-10}$ respectively.
The most 
favoured one from the
recent analyses of data \cite{Gonzalez-Garcia:2001sq} 
is the LAMSW.
In any
case, there are basically three types of neutrino spectra consistent
with the hierarchy of mass-splittings \cite{altfer}: {\em hierarchical}
($\kappa_1^2\ll\kappa_2^2\ll\kappa_3^2$), {\em ``intermediate''}
($\kappa_1^2\sim\kappa_2^2\gg\kappa_3^2$) and {\em ``degenerate''}
($\kappa_1^2\sim\kappa_2^2\sim\kappa_3^2$). 
In the usual
notation, $\Delta \kappa_{atm}^2\equiv \Delta
\kappa_{32}^2$, $\Delta \kappa_{sol}^2\equiv \Delta \kappa_{21}^2$.

Concerning the mixing angles, $\theta_{23}$ and $\theta_{13}$ are
constrained by the atmospheric and CHOOZ data to be near maximal and
minimal, respectively.  The $\theta_{12}$ angle depends on the solution
considered for the solar neutrino problem: it should be 
either near maximal (LAMSW, LOW and VO) or near  minimal (SAMSW).
Hence, the two basic
forms that $U$ can present are either a single-maximal or (more
plausibly) a bimaximal mixing matrix. Schematically,
\be
\label{Uaprox} 
U\sim\pmatrix{1 & 0 & 0\cr 0& \frac{1}{\sqrt{2}} &
\frac{1}{\sqrt{2}}\cr 0 & -\frac{1}{\sqrt{2}} & \frac{1}{\sqrt{2}} } \
\ \ \ \mathrm{or} \ \ \ \ U\sim\pmatrix{\frac{1}{\sqrt{2}} &
\frac{1}{\sqrt{2}} & 0\cr -\frac{1}{2}& \frac{1}{2}&
\frac{1}{\sqrt{2}}\cr \frac{1}{2} & -\frac{1}{2} & \frac{1}{\sqrt{2}}
}\ , \ee

Let us turn to our question of what is the most general
form of  ${\bf Y_\nu}$ and ${\bf Y_\nu^+}{\bf Y_\nu}$
compatible with all the previous  phenomenological requirements
in a see-saw scenario.
Notice, in the first place, that one can always choose to work in a basis 
of right neutrinos where ${\cal M}$ is diagonal
\be {\cal M}=\mathrm{diag}({\cal M}_1,{\cal M}_2,{\cal M}_3)\equiv
D_{\cal M}, 
\ee
with ${\cal M}_i\geq 0$. Then, from eqs.(\ref{kappa}, \ref{Udiag}),
$D_\kappa=  U^T{\bf
Y_\nu}^TD_{\sqrt{{\cal M}^{-1}}} D_{\sqrt{{\cal M}^{-1}}} {\bf Y_\nu}U$,
where, in an obvious notation, $D_{\sqrt{A}}\equiv +\sqrt{D_{A}}$.
Consequently, the most general form of ${\bf Y_\nu}$ and 
${\bf Y_\nu^+}{\bf Y_\nu}$ is
\bea
\label{Ynu}
{\bf Y_\nu}=D_{\cal \sqrt{M}} R D_{\sqrt{\kappa}} U^+  
\eea
\bea
\label{Ynu+Ynu}
{\bf Y_\nu^+}{\bf Y_\nu}= U D_{\sqrt{\kappa}} R^+ D_{\cal M} R
D_{\sqrt{\kappa}} U^+   
\eea
where $R$ is any orthogonal matrix ($R$ can be
complex provided $R^TR= {\bf 1}$).

So, besides the physical and measurable low-energy parameters, 
contained in $D_{\kappa}$ and $U$, ${\bf Y_\nu}$ and 
${\bf Y_\nu^+}{\bf Y_\nu}$ depend on the three
(unknown)  positive mass eigenvalues of the righthanded neutrinos, 
contained in $D_{\cal M}$, and
on the three  (unknown) complex parameters defining
$R$. We will see, however, that in practical cases the number  of
relevant free parameters becomes drastically reduced. 
Notice also that ${\bf Y_\nu^+}{\bf Y_\nu}$, and therefore
eq.(\ref{Ynu+Ynu}), does not depend on the ${\nu_R}$-basis, and thus on
the fact that ${\cal M}$ is diagonal or not.

Two (very) special cases of eqs.(\ref{Ynu}, \ref{Ynu+Ynu}) occur when
 $R={\bf 1}$ and when  ${\bf Y_\nu}$ 
has the form ${\bf Y_\nu}= W D_Y$, where $D_Y$
is a diagonal matrix and $W$ is a unitary matrix.
Then, there exists a basis of $L_i$, ${\nu_R}_i$ where 
all the leptonic flavour violation arises from the ${\bf Y_e}$ 
(i.e. from the charged leptons) or from 
the ${\cal M}$ (i.e. the right neutrinos) matrices respectively.
In the second case ${\bf Y_\nu^+}{\bf Y_\nu}$ is diagonal, so
the predictions for BR($l_i\rightarrow l_j, \gamma$)
are negligible.

Next, we  study the general predictions for BR($l_i\rightarrow l_j, \gamma$), 
focussing on BR($\mu \rightarrow e \gamma $),
by considering, in a separate way, some interesting scenarios that 
often appear in the literature. A more detailed discussion can be found in 
Ref.\cite{Casas:2001sr}.

\section{Predictions for BR($l_i\rightarrow l_j, \gamma$)}

\subsection*{$\nu_L$'s and $\nu_R$'s completely hierarchical}    


In this case
$D_\kappa\simeq \mathrm{diag}(0,\kappa_2,\kappa_3),\;\;\;\;  D_{\cal
{M}}\simeq \mathrm{diag}(0,0,{\cal M}_3)$.


\vspace{0.27cm}
\hspace{-0.9cm}$\star$\hspace{0.1cm}
If $R$ is a generic matrix, with $R_{32}\neq 0$ or $R_{33}\neq
0$,
$({\bf Y_\nu^+}{\bf Y_\nu})_{ij}$
is given by 
\bea
\nonumber
({\bf Y_\nu^+}{\bf Y_\nu})_{ij}\sim({\bf Y_\nu})_{3i}^*({\bf
Y_\nu})_{3j}\sim{\cal M}_3 \left[\sum_{l=2,3}
R_{3l}^*\sqrt{\kappa_l}U_{il}\right] \left[\sum_{l'=2,3}
R_{3l}\sqrt{\kappa_l'}U_{jl'}^*\right]  \eea
Parameterizing $R$ as
\bea
\label{R}
R=\pmatrix{\hat c_2\hat c_3 & -\hat c_1\hat s_3-\hat s_1\hat s_2\hat
c_3  & \hat s_1\hat s_3-\hat c_1\hat s_2\hat c_3 \cr  \hat c_2\hat s_3
& \hat c_1\hat c_3-\hat s_1\hat s_2\hat s_3   & -\hat s_1\hat c_3-\hat
c_1\hat s_2\hat s_3 \cr \hat s_2  & \hat s_1\hat c_2 & \hat c_1\hat
c_2\cr}\;,  \eea
where $\hat \theta_1$, $\hat \theta_2$, $\hat \theta_3$ are arbitrary
complex angles [eq.(\ref{R}) is sufficiently general for this case],
one obtains in particular
\bea
\label{summary3} 
({\bf Y_\nu^+}{\bf Y_\nu})_{21}\sim \frac{|{Y}_0|^2}{|\hat
s_1|^2\kappa_2+ |\hat c_1|^2\kappa_3} \hat c_1^*\hat s_1 
\sqrt{\kappa_3\kappa_2}U_{23}  U_{12}^* 
\eea
Here $|{Y}_0|^2$ is the largest eigenvalue of ${\bf Y_\nu^+}{\bf
Y_\nu}$ 
and $\hat \theta_1$ is an 
arbitrary complex angle. The branching ratio just depends
on $|{Y}_0|^2$ and $\hat \theta_1$.

For the LAMSW scenario the previous equation generically gives
BR($\mu\rightarrow e, \gamma$) above the {\em present} experimental
limits,   at least for ${Y}_0 = {\cal O}(1)$, as it occurs in the
unified scenarios. (This holds until $m_0 > 1.5$ TeV, except for a narrow 
range at low $m_0$.)
Actually, even for
${Y}_0 = {\cal O}(10^{-1})$, most of the parameter space  will be
probed in the forthcoming generation of experiments experiments.
This is illustrated in Fig.~2.

\begin{figure}[t]
\epsfxsize=15pc 
\centerline{
\epsfbox{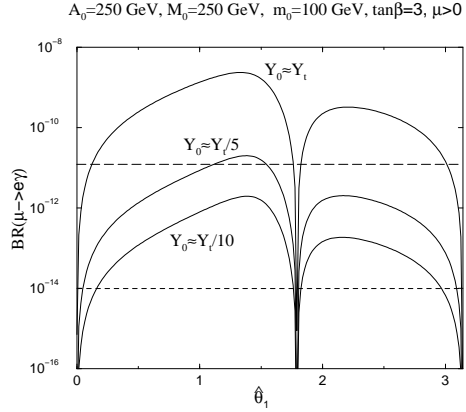} 
}
\caption{BR($\mu\rightarrow e, \gamma$) vs. the unknown angle $\hat \theta_1$
 for the case of hierarchical (left and right) neutrinos,
a typical set of 
supersymmetric parameters, and different values of the largest  
neutrino Yukawa coupling, $Y_0$, at the ``unification'' scale, 
$M_X$. $Y_t$ denotes the value of the top Yukawa coupling at that 
scale. $\hat \theta_1$ is taken real for simplicity, so the two limits
$\hat \theta_1=0,\pi$ of the horizontal axis represent the same
physical point. The dashed lines correspond to the present and forthcoming 
experimental upper bounds ${}^{15}$.
}  
\end{figure}

\vspace{0.27cm}
\hspace{-0.9cm}$\star$\hspace{0.1cm}
The only exceptions to the previous result are
  \begin{itemize}

  \item If $\hat \theta_1$ is such that $({\bf Y_\nu})_{31}\simeq 0$ or
$({\bf Y_\nu})_{32}\simeq 0$. Then $({\bf Y_\nu^+}{\bf Y_\nu})_{21}\simeq 0$
and BR($\mu\rightarrow e, \gamma$) is small. These two special values
of $\hat \theta_1$ are visible in Fig.~2 and correspond to
$\tan \hat \theta_1 \simeq
-\sqrt{\frac{\kappa_3}{\kappa_2}}\frac{V_{13}^*}{V_{12}^*} \simeq 0$
and $\tan \hat \theta_1 \simeq
-\sqrt{\frac{\kappa_3}{\kappa_2}}\frac{V_{23}^*}{V_{22}^*}$. The 
associated  textures in our basis are
\bea  {\bf Y_\nu}&\propto&\pmatrix{0 & 0 & 0\cr 0& 0& 0\cr 0 & V_{31}
&  -V_{21} }  \propto \pmatrix{0 & 0 & 0\cr 0& 0& 0\cr 0 & 1 & 1 } \ \
\ \ ,
\label{textures1} 
\\ {\bf Y_\nu}&\propto&\pmatrix{0 & 0& 0\cr 0& 0& 0\cr  -V_{31} & 0 &
V_{11} } \propto \pmatrix{0 & 0& 0\cr 0& 0& 0\cr -1 & 0 & {\sqrt{2}} }
\ \ \ \ .
\label{textures2} 
\eea
where the numerical values of the entries correspond to a  bimaximal mixing
$V$ matrix. For these cases $({\bf Y_\nu^+}{\bf Y_\nu})_{21}\propto 
{\cal M}_2$ (instead $\propto {\cal M}_3$), which can still be sizeable.
In any case other processes, as BR($\tau\rightarrow \mu, \gamma$),
are not suppressed, normally lying above the forthcoming 
experimental upper bound.

 \item If $R$ is such that ${\bf Y_\nu^+}{\bf Y_\nu}$ is
       diagonal.  This requires a very special
       form of $R$, which in particular has $R_{32},R_{33} \simeq 0$.

  \end{itemize}


\subsection*{$\nu_L$'s hierarchical and $\nu_R$'s degenerate}


In this case
$D_\kappa\simeq \mathrm{diag}(0,\kappa_2,\kappa_3),\;\;\;\;  D_{\cal
{M}}\simeq \mathrm{diag}({\cal M},{\cal M},{\cal M})$.


\vspace{0.27cm}
\hspace{-0.9cm}$\star$\hspace{0.1cm}
If $R$ is real, this scenario is very predictive. Then
 $({\bf Y_\nu^+}{\bf Y_\nu})_{ij}$ does not depend on $R$ 
\bea
\label{Ynu+Ynu23}
({\bf Y_\nu^+}{\bf Y_\nu})_{ij}=  {\cal M}\kappa_l
U_{il}U^+_{lj}\simeq   {\cal M}\left[ \kappa_2 U_{i2}U^*_{j2} +
\kappa_3 U_{i3}U^*_{j3} \right] \ ,   \eea
In particular, taking into account $U_{13}\simeq 0$,
\bea
\label{summary4}
({\bf Y_\nu^+}{\bf Y_\nu})_{21} \simeq   
{\cal M}  \kappa_2 U_{22}  U_{12}^* 
= |Y_0|^2 \frac{\kappa_2}{\kappa_3}U_{22} U_{12}^*\ ,
\eea
where $|Y_0|^2 \simeq {\cal M}\kappa_3$ is the largest eigenvalue 
of ${\bf Y_\nu^+}{\bf Y_\nu}$.

The corresponding ${\mathrm{BR}}(\mu\rightarrow e, \gamma)$ is illustrated
in  Fig.~3 for the LAMSW scenario.

\begin{figure}[t]
\epsfxsize=15pc 
\centerline{
\epsfbox{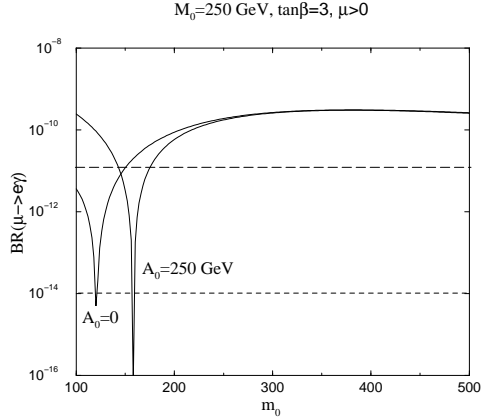} 
}
\caption{
BR($\mu\rightarrow e, \gamma$)
vs. the universal scalar mass, $m_0$, for the case of hierarchical
(degenerate) left (right) neutrinos, $R$ real and
 typical sets of 
supersymmetric parameters.
The dashed lines correspond to the present and forthcoming upper bounds.
A top-neutrino ``unification'' condition has been used to fix
the value of the largest neutrino Yukawa coupling at high energy.
The curves do not fall below the present bound until  $m_0 > 1.6$ TeV.
}
\end{figure}
The branching ratio turns out to be  already above the 
present experimental limits except for a rather small region of $m_0$ 
values which should be probed by the next generation of experiments.
Here are not special textures where the branching ratio becomes suppressed.

\vspace{0.27cm}
\hspace{-0.9cm}$\star$\hspace{0.1cm}
If $R$ is complex, the analysis is more involved since it contains
more arbitrary parameters. But 
in general the conclusion is the same:
${\mathrm{BR}}(\mu\rightarrow e, \gamma)$ is at least of the same
order as in the real case. Now there exists, however, the possibility of a
(fine-tuned) cancellation.


\subsection*{$\nu_L$'s quasi-degenerate}

%
In this case $D_\kappa\simeq \mathrm{
diag}(\kappa_1,\kappa_2,\kappa_3)$,  with
$\kappa_1\sim\kappa_2\sim\kappa_3\equiv \kappa$.  Then, it is logical
to assume that ${\cal M}$ has degenerate eigenvalues, otherwise a big
conspiracy would be  needed between ${\bf Y_\nu}$ and ${\cal M}$. Hence
$D_{\cal {M}}\simeq \mathrm{diag}({\cal M},{\cal M},{\cal M})$. 


\vspace{0.27cm}
\hspace{-0.9cm}$\star$\hspace{0.1cm}
If $R$ is real, ${\bf Y_\nu^+}{\bf Y_\nu}={\cal M} U D_{\kappa} U^+$.
In particular, since $U_{13}\simeq 0$,
\bea
\label{summary5}
({\bf Y_\nu^+}{\bf Y_\nu})_{21}={\cal M} 
\left[U_{21} U_{11}^*(\kappa_1-\kappa_2) \right]
= |Y_0|^2 U_{21}U_{11}^* \frac{\Delta \kappa_{sol}^2}{\kappa^2}
\ \ .
\eea
where, again, $|Y_0|^2 \simeq {\cal M}\kappa$ is the largest eigenvalue 
of ${\bf Y_\nu^+}{\bf Y_\nu}$. This equation is identical to
eq.(\ref{summary4}), multiplied by 
$\kappa^{-2}\sqrt{\Delta \kappa_{sol}^2 \Delta \kappa_{atm}^2}$. 
This is a factor $\sim 10^{-4}$ for the LAMSW.
Therefore all the plots representing BR($\mu\rightarrow e, \gamma$) 
in the previous scenario (Fig.~3)
are valid here, but with the vertical axis re-scaled eight 
orders of magnitude smaller. Consequently,  BR($\mu\rightarrow e, \gamma$)
is naturally suppressed below the present (and even forthcoming) limits.

\vspace{0.27cm}
\hspace{-0.9cm}$\star$\hspace{0.1cm}
If $R$ is complex, ${\bf Y_\nu^+}{\bf Y_\nu}={\cal M} U D_{\sqrt{\kappa}}R^+R
D_{\sqrt{\kappa}} U^+$, which may have sizeable off-diagonal
entries. Hence,  BR($\mu\rightarrow e, \gamma$),
could be very large in this case.

\vspace{0.27cm}
\hspace{-0.9cm}$\star$\hspace{0.1cm}
If the (quasi-) degeneracy is only partial: 
$\kappa_3\ll \kappa_1\simeq\kappa_2\equiv \kappa\sim 
\sqrt{\Delta \kappa_{atm}^2}$, 
$({\bf Y_\nu^+}{\bf Y_\nu})_{21}$
is given (for $R$ real) by 
eq.(\ref{summary4}), multiplied now by 
$\sqrt{\frac{\Delta \kappa_{sol}^2}{ \Delta \kappa_{atm}^2}}$.
This represents a suppression factor 
$\sim 10^{-1}$ for the LAMSW, which means that
 Fig.~3 should be re-scaled by a 
factor $\sim 10^{-2}$. As a consequence, 
BR($\mu\rightarrow e, \gamma$) for this partially degenerate scenario
should be testable within the next generation of experiments.
The conclusion is similar for BR($\tau\rightarrow \mu, \gamma$).

For generic complex $R$, the value of BR($\mu\rightarrow e, \gamma$)
does not get any suppression and falls naturally above the present
experimental limits.


\section{Conclusions}

If the origin of the neutrino masses is a supersymmetric see-saw, which 
is probably the most attractive scenario to explain their smallness,
then the leptonic soft breaking terms acquire off-diagonal contributions 
through the RG running, which drive non-vanishing 
BR($l_i\rightarrow l_j, \gamma$). These contributions
are proportional to $({\bf Y_\nu^+}{\bf Y_\nu})_{ij}$, where 
${\bf Y_\nu}$ is the neutrino Yukawa matrix,

Therefore, in order to make predictions for these branching ratios, one has
first to determine the most general form of 
${\bf Y_\nu}$ and ${\bf Y_\nu^+}{\bf Y_\nu}$, compatible with all 
the phenomenological requirements. This is summarized in 
eqs.(\ref{Ynu}, \ref{Ynu+Ynu}). 

Then, we have shown that the predictions for BR($\mu\rightarrow e, \gamma$) 
are normally {\em above} the {\em present} experimental limits if the three 
following conditions occur

\begin{enumerate}

\item The solution to the solar neutrino problem is the LAMSW, 
as favoured by the most recent analyses.

\item  ${Y}_0(M_X) = {\cal O}(1)$, where $|Y_0|^2$ is the largest
eigenvalue of $({\bf Y_\nu^+}{\bf Y_\nu})$. This occurs e.g. in most 
grand-unified scenarios.

\item The soft-breaking terms are generated at a high-energy scale, e.g.
$M_X$, above the Majorana mass of the right-handed neutrinos, $M$.

\end{enumerate}

\noindent
These conditions are very plausible. In our opinion, the most natural
scenarios fulfill them, but certainly there exists other
possibilities. E.g. it may happen that
supersymmetry is broken at a scale below $M$. This is the case of
gauge-mediated scenarios, where there would be no generation of
off-diagonal leptonic soft terms through the RG running.

\vspace{0.3cm}
\noindent
Even under the previous 1--3 conditions, there are physical scenarios
compatible with the present  BR($\mu\rightarrow e, \gamma$) 
experimental limits. Namely

\begin{itemize}

\item Whenever all the leptonic flavour violation can be attributed to
the sector of right-handed neutrinos.  
In this case there is no RG generation of non-diagonal
soft terms.

\item In the scenario of  hierarchical (left and right) 
 neutrino masses, {\em if} ${\bf Y_\nu}$ has (in our basis)
one of the two special
textures shown in eqs.(\ref{textures1}, \ref{textures2}).

\item If the left-handed neutrinos are quasi-degenerate and 
the $R$ matrix in eq.(\ref{Ynu}) is real.

\end{itemize}

In our opinion, the scenario 
of quasi-degenerate neutrinos and the one with gauge mediated supersymmetry 
breaking  represent the most plausible explanations
to the absence of $\mu\rightarrow e, \gamma$ observations, specially if 
the absence persists after the next generation of experiments.

As a final conclusion, the discovery of neutrino oscillations makes
much more plausible the possibility of observing
lepton-flavour-violation processes, specially $\mu\rightarrow e,
\gamma$, if the theory is supersymmetric and  the neutrino masses are
generated by a see-saw mechanism.  Large regions of the parameter
space are {\em already excluded} on these grounds, and there exists great
chances to observe $\mu\rightarrow e, \gamma$ in the near future (PSI,
2003). This means that, hopefully, we will have signals of
supersymmetry before LHC.

\end{document}